\documentstyle[twocolumn,aps,graphics]{revtex}

\begin{document}

\title{Evolution of an Atom Impeded by Measurement: The Quantum Zeno Effect}

\author{Chr. Wunderlich, Chr. Balzer, and P.E. Toschek}

\address{Institut f\"{u}r Laser-Physik, Universit\"{a}t Hamburg, 
Jungiusstr. 9, D-20355 Hamburg, Germany}

\date{February 6, 2001}
\maketitle


\begin{abstract}
A quantum system being observed evolves more slowly. This 
`'quantum Zeno effect'' is reviewed with respect to a previous 
attempt of demonstration, and to subsequent criticism of the 
significance of the findings. A recent experiment on an {\it 
individual} cold trapped ion has been capable of revealing the 
micro-state of this quantum system, such that the effect of 
measurement is indeed discriminated from dephasing of the quantum 
state by either the meter or the environment.

\end{abstract}



\section{Introduction}

Quantum mechanics is a statistical theory whose elements are 
ensembles of quantum systems, or ensembles of measurements on the 
same quantum system. The quantities to be compared with 
observation are expectation values, weighted averages over 
eigenvalues that are the hypothetical results of measurements on 
an individual system. Small wonder that no particular 
significance has been attributed to such individual measurements, 
and this neglect was vindicated the more as Erwin Schr\"{o}dinger 
commented in 1952:
\begin{quote}
'' ... this is the obvious way of registering the fact, that we \textit{%
never }experiment with just \textit{one} electron or atom or 
(small) molecule. In thought-experiments we sometimes assume that 
we do; this invariably entails ridiculous consequences ... In the 
first place it is fair to state that we are not 
\textit{experimenting} with single particles, any more than we 
can raise Ichthyosauria in the zoo. We are scrutinising records 
of events long after they have happened ... '' [1].
\end{quote}

However, less than 30 years after this verdict we have learnt to 
confine individual ions in an electrodynamic trap, to laser-cool 
their vibrational motion, and to store them for hours or even 
days [2]. Since then, repeated measurements on an individual and 
identically prepared quantum system are indeed available. Not 
only that such a system may serve as an ultra-precise frequency 
standard [3] or as a building block for logical gates in quantum 
information processing [4]: The entire concept of individual 
measurements on such a system deserves reconsideration [5,6].
Here, unlike with an \textit{ensemble} of quantum systems, the \textit{%
micro-state} of the individual system is revealed with a 
measurement, and from the eigenvalues that show up as the 
results, correlations may be derived up to an order only limited 
by the number of sequential measurements in a time series. For an 
example, the correlations of two subsequent measurements 
resulting in different eigenvalues --- one of which might be
represented by a null signal of the meter --- would characterize \textit{%
transition} processes, as the obervation of excitation or 
deexcitation of the quantum system. On the other hand, the 
correlation of a series of \textit{equal} results --- even when 
represented by null signals --- characterizes the system as 
effectively invariable under the reiterated procedure of 
attempted preparation and probing. With such an individual 
quantum system at hand, it seems tempting to reexamine its 
evolution under quasi-continuous probing, implemented by making 
increase the number of measurements. The quantum system, prepared 
in a particular eigenstate of energy and evolving freely, shows 
its transition probability varying with the square of lapsed 
time, $\left( \Delta t\right) ^{2}.$ Each one of a series of $N$ 
measurements equally distributed over the time interval $T$ 
resets the evolution with high probability, if $\Delta t$ is 
small enough. Then, the overall transition probability after $T$ 
is $N\left( \Delta t\right) ^{2}=T^{2}/N$ which vanishes for 
large enough $N$ [7]. This impeded evolution is tantamount to the 
quantum Zeno effect (QZE), or Zeno ''paradox'' [8-10]. It is of 
utmost importance for the characterization and appreciation of 
the relationship of the quantum system with its classical 
measuring apparatus [11].

\section{What does Quantum Zeno signify?}

A previous attempt of demonstrating the inhibition of quantum 
evolution has made use of a large number of electromagnetically 
trapped ions, whose ground-state hyperfine structure well 
represents a two-level system, interacting with a resonant radio 
wave, for the coherent drive. The sample of ions was irradiated, 
during the driving pulse, by $N$ resonant light pulses, applied 
as the perturbation [12]. Perfect agreement with the predictions 
of quantum mechanics has been achieved. However, the subsequent 
discussion has put into question the relevance of the findings to 
the demonstration of the evolution of the quantum system impeded 
by\textit{\ measurement}, and it has identified requirements for 
such an unequivocal demonstration of the QZE:

  (i) Mere interventions with the quantum system --- as, 
for example, the irradiation with resonant light pulses --- do 
not qualify as measurements, since some of the results of these 
interventions may cancel and not show up in the final evaluation 
of the transition probability. Instead of the ''net'' transition 
probability, evaluated in such a series of interventions with 
final probing, the true probability must be measured that 
includes the results of each of the interventions, in order to 
exclude the falsifying effect of back-and-forth transitional acts 
[13]. By the same token, the state of an\textit{\ individual} 
quantum system should be detected after each of the interventions 
in order to exclude compensating transitions in different members 
of a collective system.

  (ii) Possible reaction of the meter and/or the 
environment upon the phase of the quantum object's wave function 
is supposed to mimic the effect of measurement, \textit{except 
}with an individual quantum system [14-16]. Such a system quite 
generally seems to render the only chance for discrimination 
between physical dephasing, and the effect of entanglement of 
quantum object and meter that emerges with a measurement. Note 
that the destruction of an ensemble's coherence may modify the 
temporal evolution irrespective of any correlation with the 
meter. Moreover, Spiller has shown that different types of 
microscopic dynamics may cause identical averaged results even 
though individual realizations are very different [15]. In 
particular, he has shown that, for example, the decohering random 
precession of spins produces --- at the statistical level --- the 
same results as projective measurements.

  (iii) Even with the use of a single quantum system, 
simulation of the effect of measurements by physical phase 
perturbations upon this quantum object can and should be 
disproved. A general and sufficient approach to this 
characterization --- although not a necessary one --- is to 
consider measurements that lack a signal of the meter, and that 
have been called ''negative-result'' measurements [11]. However, 
any measurement will do whose reaction upon the quantum object 
demonstrably does not qualify for feigning the effect of the 
measurement. In particular, quantum non-demolition measurements 
[14] comply with this requirement.

  The above preconditions, as well as the relevance of 
state reduction and projection postulate to the Zeno problem, 
have been detailed elsewhere [17].

  Recently, an experiment on an \textit{individual} 
atomic two-level
system has included read-out of the system's state \textit{after each probing%
}. It has resulted in proving that the effect of the 
interventions by the repeated measurements is due to 
entanglement, not to dynamic interaction [18]. This proof seems 
to give no leeway anymore for alternative interpretation.

\section{The experiment}

  The experimental implementation of an individual 
quantum system subjected to repeated intervention and probing 
preferentially includes an atom, represented as a two-level 
system with little relaxation whose relevant levels are labelled 
0 and 1 (Fig. 1). A field of radiation coherently drives the 
resonance for a preselected time interval $\tau $, such that 
transitions take place with flopping frequency $\Omega =\theta 
/\tau $. Here, $\theta $ is the radiative pulse area or 
''nutation angle'', and cos $\theta $ determines the energy of 
the coherent atomic superposition state that is prepared by the 
pulse. Probe-light pulses of length $\tau _{p}$ alternate with 
the reiterated preparations; they elicit resonance light 
scattering, if the atom is found in state 0, but they don't, if 
the atom has been taken, by the coherent drive, to level 1 [19]. 
Measurements that include probing results of the latter kind lack 
the observation of scattered light, i.e., they are signal-less 
and of the ''negative-result'' type.

  The individual quantum system is represented by a 
$^{172}$Yb$^{+}$ ion confined in a 1-mm-sized electrodynamic trap 
and cooled well into the Lamb-Dicke regime (i.e., motional 
excursion $\ll $ light wavelength). State 0 is identified with 
the S$_{1/2}$ ground state, and state 1 with the metastable state 
D$_{5/2}$ connected with the ground state by an electric 
quadrupole transition that is coherently driven by 411-nm light 
of high monochromaticity and coherence. Light at 369nm excites 
resonance scattering on the electric dipole resonance 0-2, with 2 
representing the P$_{1/2}$ state, if the ion resides in its 
ground state. The resonantly scattered light serves as the probe 
signal, whose values ''on'' and ''off'' being correlated with the 
ion found in its S$_{1/2}$ and D$_{5/2}$ states, respectively. 
Spurious optical pumping to the $^{2}$F$_{7/2}$ level by the 
drive light and to the $^{2}$D$_{3/2}$ level by the probe light is 
eliminated by saturated repumping to the ground state, using 
light at 638nm and 609nm, respectively. A pair of drive and probe 
pulses make a measurement, and 500 measurements a trajectory of 
the ion's evolution. A section of such a trajectory, with a 
particular degree of excitation characterized by a nutational 
phase $\theta =\Omega \tau $ of the drive pulse, is shown in Fig. 
2. Note that a pair of measurements with the result ''on'' and 
subsequent ''off'' (''off'', ''on'') signifies an act of 
excitation (deexcitation), separated by sequences of identical 
results. These sequences indicate the ion to be ''set back'' to 
its initial state by the probing after each coherent preparation 
in a superposition state characterized by $\theta $ --- or, 
alternatively that the preparations are frustrated, and nothing 
has happened.

  This state of affairs is substantiated by the recording 
of an excitation spectrum, when the drive light is stepwise 
scanned across the E2 resonance (Fig. 3). The ion's transition 
probability oscillates upon scanning the detuning $\Delta =\omega 
-\omega _{0}$ and the effective phase angle $\theta 
_{eff}=\sqrt{\Omega ^{2}+\Delta ^{2}}\tau $. It also assumes peak 
values well above 1/2, when the Zeeman splitting of the ground 
state is taken into account. Both these features characterize as 
coherent the interaction with the drive light. In fact, the 
spectral oscillations amount to a stroboscopic recording of the 
ion's Rabi nutation, as it is simulated in the bottom part of 
Fig. 3.

  The precondition of coherent driving being confirmed, 
we analyze the ion's temporal evolution using the distribution of 
the sequences of \textit{equal} results. Let us tentatively 
assume that the acts of frustrated probing that render zero 
signals do not interfere with the intertwined coherent driving. 
The conditional probability for the ion to be found in its 
initial eigenstate would write $V_{coh}(q)=\cos ^{2}(q\Omega \tau 
/2),$ where $q$ is the number of identical results in the 
sequence. This distribution of probability is shown in Fig. 4: 
The probability completely vanishes for odd multiples of $\theta 
=\pi $ pulses. In contrast, the coherent evolution set back by 
the probing --- even with a signal-less
result --- leaves that conditional probability as $V(q)=p^{q}$, where $%
p_{0}= $ $p_{1}=p=\cos ^{2}(\theta /2)$, and $p_{i}$ is the 
probability of finding ion in state $i$.

  From the observed trajectories, the statistics of the 
sequences of equal results allows us to derive the likeliness of 
the ion remaining in its state. These sequences represent a 
sub-ensemble of the general correlations mentioned above. With 
$U(q)$ being the number of sequences of q equal results, 
$V_{obs}(q-1)=U(q)/U(1).$ Note that for a quantitative evaluation, 
energy and phase relaxation must be included in the model [18]. 
This modification makes $p_{0}$ and $p_{1}$ disagree.

  The statistics of the sequences is shown in Fig. 5 for 
three different values of $\theta _{eff}$, and for both the light 
scattering ''on'' and ''off''. The data show the exponential 
distribution of $V(q-1)$ and demonstrate the effect of the 
measurement, i.e., the non-local correlation of quantum system 
and meter [11].

  Although we have restricted the discussion to the 
null-signal measurements, a thorough analysis shows that even the 
distributions of sequences of ''on'' results prove the effect of 
the measurements to be responsible for the exponential 
distributions: The reaction of the meter (or the environment) is 
as small as not to allow interpretation as a dynamical effect 
[17,18].

\section{Conclusions}

  The retardation or even inhibition of the evolution of 
a quantum system has been reconsidered with respect to three 
items:

  (i) A previous experiment that has made use, for the 
quantum system, of a large ensemble of ions. (ii) Some of the 
subsequent criticism
whose central argument refers to the effect of measurement on an \textit{%
ensemble} being indistinguishable from a dynamical, reactive 
effect of the meter and/or the environment.

  (iii) A recent experiment on an \textit{individual} 
ion, that was, moreover, designed to register the relevant state 
of the quantum system after \textit{everyone }of the radiative 
interventions by the probe.

  Serious objections had been raised on the nature of the 
impeding effect: In order to convincingly attribute the effect to 
the measurement, i.e., for a signature of entanglement of the 
quantum object and its meter, the measurements had been required 
to be non-local and of the negative-result type. The single-ion 
measurements with ''off'' results comply with this requirement. 
Moreover, even the ''on'' results can be demonstrated not to 
arise from the reactive effect of the interventions by the probe 
light [17,18].

  Now that the QZE --- and even the corresponding 
''paradox'' [11]
--- has been demonstrated unambiguously, the pitfalls of interpretation are
eliminated. QZE seems to characterize the non-local correlation 
of a quantum system with a macroscopic meter in a way that is 
much alike the violation of Bell's inequalities characterizing 
the non-local correlation of two quantum systems being 
macroscopically separated [20].

  This work was supported by the K\"{o}rber-Stiftung, by 
the Hamburgische Wissenschaftliche Stiftung, and by the 
ZEIT-Stiftung, Hamburg.

\section*{References}

\begin{enumerate}
\item  E. Schr\"{o}dinger, Brit.J.Phil.Sci. \textbf{3}, 109 (1952).
\item  W. Neuhauser, M. Hohenstatt, P.E.\ Toschek, and H.G. Dehmelt,
Phys.Rev. A \textbf{22}, 1137 (1980).
\item  e.g., A. De Marchi, ed., \textit{Freqeuncy Standards and Metrology},
Springer-Verlag, Berlin, 1988.
\item  e.g., A. Steane,
Appl.Phys. B \textbf{64}, 623 (1997).
\item  B. Appasamy, Y. Stalgies, and P.E. Toschek, Phys.Rev.Lett. \textbf{80}%
, 2805 (1998).
\item  R. Huesmann, Ch. Balzer, Ph. Courteille, W. Neuhauser, and P.E.
Toschek, Phys.Rev.Lett.\textbf{\ 82}, 1611 (1999).
\item  A. Beige and G.C. Hegerfeldt, Phys.Rev. A \textbf{53}, 53 (1996).
\item  L.A. Khalfin, Pis'ma Zh. Eksp. Teor. Fiz. \textbf{8}, 106 (1968);
[JETP Lett.\textbf{\ 8}, 65 (1968)].
\item  L. Fonda, G.C. Ghirardi, A. Rimini, and T. Weber, Nuovo Cimento A 
\textbf{15}, 689 (1973).
\item  B. Misra and E.C.G. Sudarshan, J.Math.Phys. (N.Y.) \textbf{18}, 756
(1977).
\item  D. Home and M.A.B. Whitaker, Ann.Phys. N.Y. \textbf{258}, 237 (1997).
\item  W.M. Itano, D.J. Heinzen, J.J. Bollinger, and D.J. Wineland,
Phys.Rev. A \textbf{41}, 2295 (1990); \textit{loc.cit}. \thinspace \ \textbf{%
43}, 5168 (1991).
\item  H. Nakazato, M. Namiki, S. Pascazio, and H. Rauch, Phys.Lett. A 
\textbf{217}, 203 (1996).
\item  V.B. Braginsky and F.Ya. Khalili, \textit{Quantum Measurement},
Cambridge University Press, Cambridge, MA, 1992.
\item  T.P. Spiller, Phys.Lett. A \textbf{192}, 163 (1994).
\item  O. Alter and Y. Yamamoto, Phys.Rev. A \textbf{55}, R2499 (1997).
\item  P.E. Toschek, and Ch. Wunderlich, Eur. Phys. J.
D {\bf 14}, 387 (2001).
\item  Chr. Balzer, R. Huesmann, W. Neuhauser, and P.E.\ Toschek, Opt.Comm. 
\textbf{180}, 115 (2000).
\item  B. Appasamy, I. Siemers, Y. Stalgies, J. Eschner, R. Blatt, W.
Neuhauser, and P.E. Toschek, Appl.Phys. B \textbf{60}, 473 (1995).
\item  A. Aspect, J. Dalibard, and G. Rogers, Phys.Rev.Lett. \textbf{49},
1804 (1982). 
\end{enumerate}

\newpage
\section*{Figure Captions}

\medskip

Fig. 1: Level scheme of an ion $(Yb^{+};$level 0$:S_{1/2},$1: $D_{5/2},$2: $%
P_{1/2})$ interacting with monochromatic drive light resonant 
with $E2$ transition, and with probe light that excites light 
scattering on a resonance transition.

Fig. 2: Part of trajectory of results of measurements each of 
which consists of a drive and a probe pulse applied to the ion.

Fig. 3: Probability of excitation, vs detuning $\nu -$ $\nu _{0}$ 
by 20kHz steps of the drive (top). Note the first-order vibronic 
sideband at 1.3 MHz. Within a small range close to resonance, 
detuning replaces variation of the drive-pulse length $\tau $ . 
The spectrum of absorption is superimposed by stroboscopic 
sampling of the ion's Rabi nutation, as demonstrated by a 
simulation with small steps on an expanded scale (bottom).

Fig. 4: Probability of ion remaining in initial state. $%
V_{coh}(q)=\cos ^{2}(q\theta /2)$: ion unobserved, unaffected by 
measurements. $V(q)=\cos ^{2q}(\theta /2)$: ion observed; $q$ is 
the length of sequence. In the graph, $\theta =2.$

Fig. 5: Probability $U(q)/U(1)$ of uninterrupted sequences of $q$ 
''on'' results (white dots) and ''off'' results (black dots). The 
lines how the distributions of probabilities $V(q-1)$ for the 
ion's evolution on its drive transition, according to 
 $V(q-1)=p_i^{q-1}$, with relaxation included. Here, 
 $i=0,1$, $p_i=1-f_i B_i(1-e^{a+b}\cos\theta)$,
 $B_0=(\Omega^2/2)/(\Omega^2+\Gamma\gamma)$,
 $B_1=1-B_0$, $2a=\gamma\tau=\gamma_{ph}\tau+(\Gamma/2)\tau$,
 $2b=\gamma\tau$,
 $\theta^2=(\Omega\tau)^2-(a-b)^2$, $\Omega$ and $\Gamma$
 are Rabi frequency, and decay rate of inversion, respectively, and
$\gamma_{ph}=(2a-b)/\tau$ is the rate of phase diffusion of the 
drive light. In addition, the finite length of the trajectories 
of measurements has been taken into account.  $\theta$ and 
$f_{1}$ from fit; values $f_{1}<1$ indicate redistribution, over 
sublevels, by cycles of spontaeous decay and reexcitation.

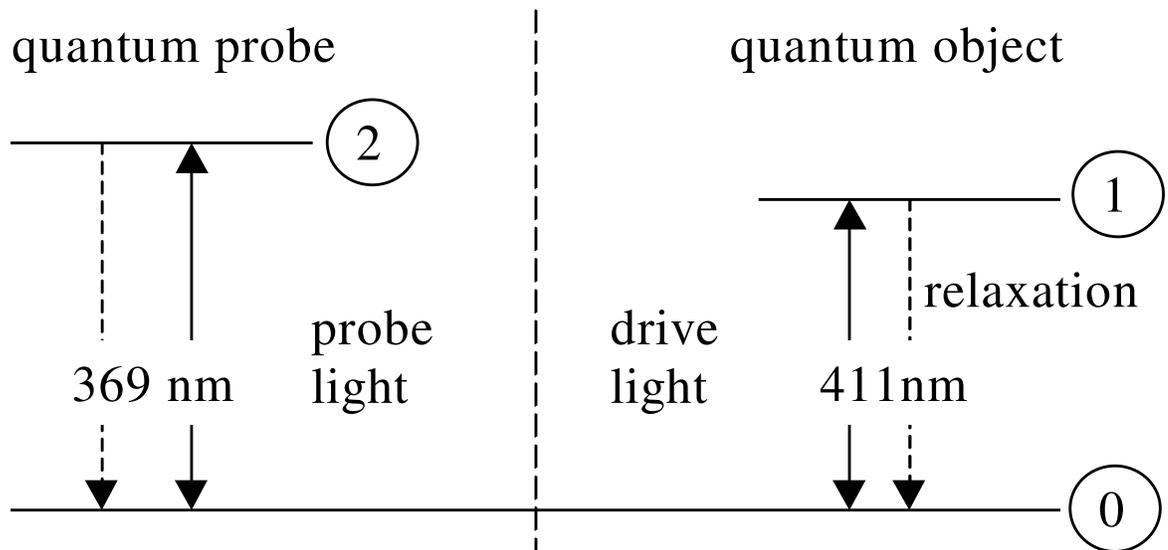
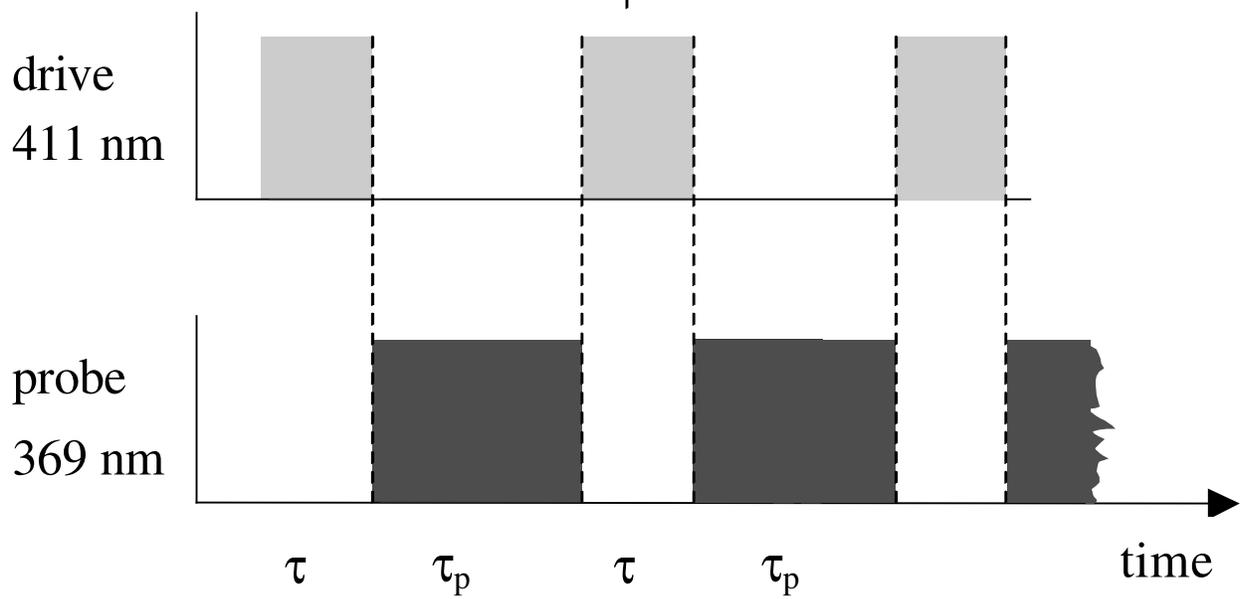
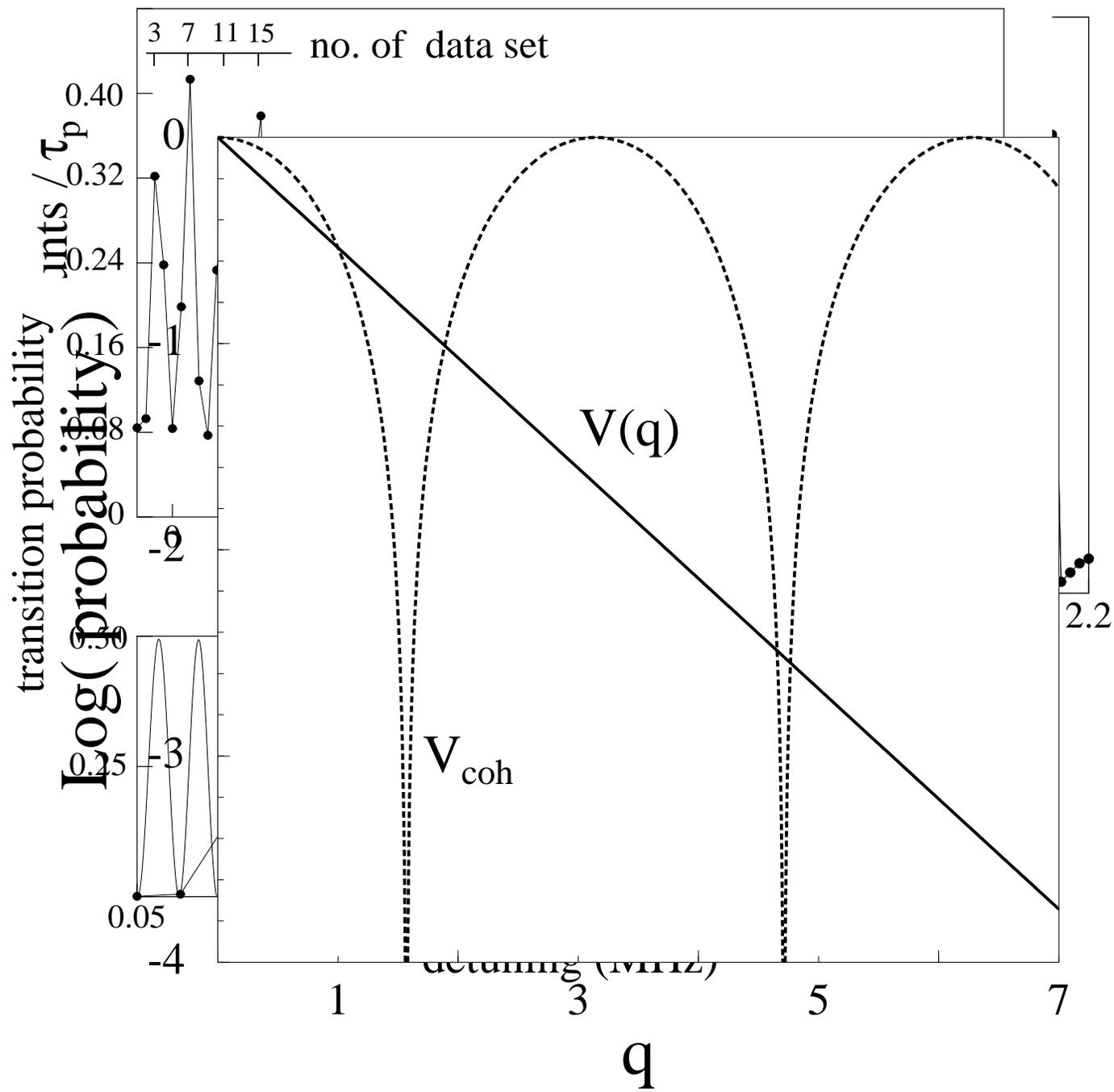
\end{document}